\documentclass[manuscript]{acmart}
\usepackage{graphicx}
\graphicspath{ {./images/} }

\AtBeginDocument{%
  \providecommand\BibTeX{{%
    \normalfont B\kern-0.5em{\scshape i\kern-0.25em b}\kern-0.8em\TeX}}}

\copyrightyear{2021}
\acmYear{2021}
\setcopyright{acmcopyright}\acmConference[IUI '21]{26th International Conference on Intelligent User Interfaces}{April 14--17, 2021}{College Station, TX, USA}
\acmBooktitle{26th International Conference on Intelligent User Interfaces (IUI '21), April 14--17, 2021, College Station, TX, USA}
\acmPrice{15.00}
\acmDOI{10.1145/3397481.3450698}
\acmISBN{978-1-4503-8017-1/21/04}

\begin{document}

\title[Increasing the Speed and Accuracy of Data Labeling Through an AI Assisted Interface]{Increasing the Speed and Accuracy of Data Labeling \\Through an AI Assisted Interface}


\author{Michael Desmond}
\email{mdesmond@us.ibm.com}


\author{Zahra Ashktorab}
\email{Zahra.Ashktorab1@ibm.com}

\author{Michelle Brachman}
\email{michelle.brachman@ibm.com}

\author{Kristina Brimijoin}
\email{kbrimij@us.ibm.com}

\author{Evelyn Duesterwald}
\email{duester@us.ibm.com}

\author{Casey Dugan}
\email{cadugan@us.ibm.com}

\author{Catherine Finegan-Dollak}
\email{cfd@ibm.com}

\author{Michael Muller}
\email{michael_muller@us.ibm.com}

\author{Narendra Nath Joshi}
\email{Narendra.Nath.Joshi@ibm.com}

\author{Qian Pan}
\email{Qian.Pan@ibm.com}

\author{Aabhas Sharma}
\email{Aabhas.Sharma@ibm.com}


\affiliation{%
  \institution{IBM Research}
  \country{USA}
}

\authorsaddresses{%
Authors' addresses: Michael Desmond, mdesmond@us.ibm.com; Zahra Ashktorab, Zahra.Ashktorab1@ibm.com; Kristina Brimijoin, kbrimij@us.ibm.com; Evelyn Duesterwald, duester@us.ibm.com; and Catherine Finegan-
Dollak, cfd@ibm.com: IBM Research Yorktown, 1101 Kitchawan Rd., Yorktown Heights, New York, USA.
Michelle Brachman, michelle.brachman@
ibm.com; Casey Dugan, cadugan@us.ibm.com; 
Michael Muller, michael\_muller@us.ibm.com; 
Narendra Nath Joshi, Narendra.Nath.Joshi@ibm.com; Qian Pan, Qian.Pan@ibm.com;
and Aabhas Sharma, Aabhas.Sharma@ibm.com: IBM Research Cambridge, 75 Binney St., Cambridge, Massachusetts, USA.
}

\renewcommand{\shortauthors}{Desmond, et al.}

\begin{abstract}

  Labeling data is an important step in the supervised machine learning lifecycle. 
  It is a laborious human activity comprised of repeated decision making: the human labeler decides which of several potential labels to apply to each example. Prior work has shown that providing AI assistance can improve the accuracy of binary decision tasks. However, the role of AI assistance in more complex data-labeling scenarios with a larger set of labels has not yet been explored. We designed an AI labeling assistant that uses a semi-supervised learning algorithm to predict the most probable labels for each example. We leverage these predictions to provide assistance in two ways: (i) providing a label recommendation and (ii) reducing the labeler's decision space by focusing their attention on only the most probable labels. 
  We conducted a user study (n=54) to evaluate an AI-assisted interface for data labeling in this context. Our results highlight that the AI assistance improves both labeler accuracy and speed,  especially when the labeler finds the correct label in the reduced label space. We  discuss findings related to the presentation of AI assistance and design implications for intelligent labeling interfaces. 

\end{abstract}




\maketitle

\section{Introduction}

As AI becomes ever more ubiquitous, so too does the need to label and annotate training data. Despite significant algorithmic advances \cite{vinyals2016matching, arazo2020pseudo} and recent labeling paradigm shifts \cite{ratner2017snorkel}, human-in-the-loop data labeling still remains the most reliable method of acquiring labeled data. But humans make mistakes, and human time is costly. 
Many real-world AI systems, such as conversational agents, require labelers to work with large label sets containing tens, hundreds, 
or even thousands 
of labels. As the number of labels increases, the labeler's decision process becomes more complex, and the mechanics of labeling become more laborious and time consuming.

A common paradigm for supporting and optimizing human decision processes is to leverage human-AI collaboration \cite{cai2019hello,wang2019human,wolfblomberg2019}. Lai and Tan \cite{lai2019human} have shown that AI assistance---in the form of exposing a human to the predictions of an AI model---can significantly improve the accuracy of human decisions on a dichotomous task. This finding is of particular interest for data labeling, a process 
of repeatedly capturing and encoding human decisions.
We show that the benefits of AI assistance for decision making can also be leveraged in more complex data labeling scenarios with large label sets.

We designed and evaluated an \textit{AI assisted data-labeling} interface wherein a predictive algorithm assists the human labeler by (a) identifying the most probable labels that apply to the unlabeled example and (b) reducing the labeling \textit{decision space} to help the labeler quickly choose the correct label. 
We conducted a user study, recruiting 54 participants from Amazon Mechanical Turk to perform a series of labeling tasks with a label set of 21 labels. One group of participants, the baseline, were unassisted, while the other two groups were assisted by two variants of an AI model, a \textit{weak} minimally trained model and a \textit{strong} model.  The \textit{strong} model had access to more labeled data and thus exhibited better predictive performance (accuracy) than the weak model. Both AI models were trained using a transductive semi-supervised learning algorithm \cite{zhou2004learning}. Labeling assistance was provided by using the model's calculated label probability distribution to rank, annotate and reduce
the set of labels presented to the labeler. The result is a set of recommended labels 
to focus the labeler's attention on the most likely label choices.
Our experiment set out to test the following hypotheses:
\begin{itemize}
\item 
\textbf{H1:} Providing AI assistance \textbf{increases human accuracy} during data-labeling tasks.
\item
\textbf{H2:} Providing AI assistance \textbf{decreases the time required} to perform data-labeling tasks.
\item 
\textbf{H3:} Improving the accuracy of AI assistance improves the accuracy of the human labeler.
\end{itemize}

Our results indicate that human labelers are significantly more accurate when assisted by a predictive model during labeling activities. Overall our AI assisted participants were ~6\% more accurate than the unassisted participants. We found no significant difference in accuracy between weaker (less accurate) AI assistance and stronger (more accurate) AI assistance. This was somewhat surprising and suggests that labelers are unlikely to overly rely on assistance; instead, they can discern when the predictions are inaccurate and make their own labeling decisions.  We also found  significant improvement in the amount of time required to perform labeling tasks in the presence of AI assistance. Labeling speed improvements were especially pronounced when labelers found the desired label in the reduced label space provided by the AI assistant. 
This paper makes the following contributions:

\begin{itemize}
   \item We developed an AI-assisted data-labeling approach where, for each example presented to the labeler, a predictive model ranks, annotates, and reduces the label set, focusing the labeler's attention on the most probable labels.
\item We conducted a user study with 54 participants from Mechanical Turk to evaluate the impact of our AI assistance on human labeler performance demonstrating that AI assistance significantly improves labeler accuracy. 
\item The user study also shows that AI assistance reduces the time taken to complete a  labeling task, especially when labelers find a label among the reduced set of labels recommended. 
\item We discuss the implications of our findings around the design of AI assistance for labeling tasks. 
\end{itemize}

\section{Related Work}\label{sec:related}

\textbf{Human-AI Interaction}:
Human-computer interaction is increasingly concerned with the collaboration between AI and human \cite{stinson2018healthy,cai2019human}.  
Users are optimistic about the incorporation of AI-assisted decision making in data science \cite{wang2019human} and pedagogical tools \cite{abu2018intelligent,stinson2018healthy} and researchers are interested in the outcome of these interactions
\cite{oh2018lead, wang2019human}.
Joint human-AI decision making outcomes can be improved if user trust towards the AI  is calibrated appropriately \cite{zhang2020effect} and when there is an appropriate level of contribution from each party in a human-AI collaborative context \cite{lai2019human,mackeprang2019discovering}.

 An implicit part of AI-assisted decision making is how the AI system's results are presented to the user. 
 Factors like explainability and transparency can influence humans' impressions and trust in an AI system.
 Studies investigating explainability have explored the role of transparency in AI systems. The Expectation Confirmation Model (ECM) states that user expectations are directly related to the user satisfaction of a system \cite{bhattacherjee2001understanding}. 
One way to set user expectation is by making the system more transparent by showing accuracy or confidence of a model. 
Researchers have investigated making the AI system more transparent to users by displaying model predictions and confidence values \cite{rosenthal2010towards,lai2019human}. 
Rosenthal and Dey~\cite{rosenthal2010towards} explored the types of information that are most useful to present to labelers to maximize labeling accuracy. They found that showing model predictions and uncertainty had a positive effect on accuracy.
Lai and Tan.  \cite{lai2019human} demonstrated that providing humans with machine predictions significantly improved human decision-making performance in a 
deception-detection task. They found that showing predictions of an AI model resulted in a 21\% accuracy improvement and showing the predictions along with confidence scores resulted in a 46\% relative improvement. 
In a similar study, Zhang et al. \cite{zhang2020effect} measured improvement in users' trust when provided with prediction and confidence scores but found no significant improvement in accuracy. The authors attributed the insignificant accuracy gain to the human and AI having little performance divergence on the task. The human and the AI rarely disagreed. 

In the aforementioned studies, the human subjects were presented with dichotomous tasks. 
Our aim is to evaluate accuracy and efficiency effects of AI assistance but in a more complex decision making scenarios, such as choosing amongst a large label set.




\textbf{Labeling Assistance Algorithms \& Tools}:
Since labeling data is considered time-consuming, expensive and difficult, a significant body of work has focused on ways to increase the labeler's accuracy and efficiency, and reduce tedium. Many of these are interactive machine learning approaches \cite{dudley2018review} in which human judgement and computational intelligence are paired in a cooperative setting. 

Active Learning \cite{settles2009active} is a labeling paradigm in which the learner (the model being trained or a proxy) assists the human labeler by choosing which examples to label from a pool of unlabeled data. Active learning is a corner stone of interactive machine learning and has been shown to significantly reduce the amount of data that needs to be labeled; however, it does not make individual labeling decisions easier, and may make labeling more difficult due to use of uncertainty as a common selection heuristic ~\cite{Settles2008}.

Sun et al. \cite{sun2017label} studied visualizations of the learning process during interactive labeling, and found that visual representations of model performance, such as model predictions, improved understanding and motivation and helped to avoid redundant labeling. Rosenthal and Dey \cite{rosenthal2010towards} also found that model predictions and uncertainty scores helped to improve labeling accuracy. For instance AILA \cite{choi2019aila} incorporates label predictions in an assisted labeling system.

A particular aspect of labeling accuracy is related to consistency. 
Kulesza et al. \cite{kulesza2014structured} introduced \textit{concept evolution}, a phenomenon where the labeler's understanding of a label changes over time due to exposure to data, resulting in temporally inconsistent labels. It's reasonable to assume that the predictions of a model might help to improve accuracy by recommending labels in a manner consistent with existing labeled examples. 

A variety of more elaborate assisted labeling systems have been proposed. 
Settles \cite{settles2011closing} designed a dual interface labeling tool entitled ``Dualist'' that uses an information gain algorithm to learn feature-label associations from labeled data. The associations are presented to the labeler who can then confirm the association, thus rapidly labeling the data in a very efficient manner. Studies indicated that Dualist significantly improved labeling efficiency. In a similar vein Williams et al. \cite{williams2015rapidly} developed an interactive approach to intent labeling for text-based virtual assistants. An initial search of an unlabeled corpora provides training data for a machine learning model, which is then applied to the unlabeled corpora in an active learning style refinement process. The labeler's experience is simplified to a sequence of binary labeling tasks performed with the aid of model predictions. Ratner et al. \cite{ratner2020snorkel} introduced the idea of \textit{weak supervision}, an approach to data labeling where human provided heuristics (labeling functions) are used in conjunction with a generative model to label data at scale.
These projects show that the application of predictive models has potential to transform the labeling process into a simpler and less tedious experience. 

\textbf{Overreliance}:
``The tendency to over-rely'' or automation bias \cite{goddard2012automation} 
has been investigated in areas such as aviation \cite{riley1994human}, as well as medicine\cite{goddard2012automation}. In human-AI collaborative environments, humans 
make decisions based in part on the AI's recommendations. When humans overly rely, they are not using their own 
cognitive
capacities to make a decision and rather are blindly following the system's recommendations. Any ``human in the loop'' process relies on humans to provide valuable feedback.
Thus, overreliance, \textit{i.e.}, the human not using their judgment to make a decision, can be detrimental to that process \cite{shneiderman2020human}.
Labelers overly relying on AI Assistance is a concern. We know that labeling tasks are tedious, and the temptation to just leave the decision to the predictive model is common. We study overreliance by considering how humans perform when working with two different AI systems, one more accurate than the other. 


\section{AI-Assisted Labeling}

A driving goal of our work was to understand how to infuse AI assistance into the labeling decision process to enhance human performance.  In particular, we wanted to study how the predictions of an AI model can be used to recommend labels and reduce the decision space that the labeler needs to consider. To further this work we implemented an AI Assisted Labeling system. 
The system is driven by a semi-supervised learning algorithm that uses both labeled and unlabeled data to generate label probability distributions for unlabeled data. The label distributions are used to both rank and annotate the labels that are presented to the labeler. The system interface presents only the five most probable labels
(referred to as the "top 5"), but the user has the option to view the rest of the labels, if necessary. 



\subsection{AI Assistance Algorithm}
At the core of our AI Assisted Labeling system is \textit{label spreading} \cite{zhou2004learning}, a transductive semi-supervised learning algorithm that predicts label distributions by propagating label information probabilistically via a graph representation of both labeled and unlabeled data. The algorithm is \textit{transductive}, meaning that it predicts labels for the unlabeled examples that have already been seen by the algorithm, as opposed to a generalizing model that predicts labels for unseen examples. 
In general, semi-supervised learning \cite{van2020survey} is an good choice for implementing labeling assistance due to the coexistence of labeled and unlabeled data. Moreover, transductive graph-based approaches are suitable as they can still operate reliably even if the amount of labeled data is limited. Our labeling system invokes label spreading after a small number of seed labels have been obtained in order to create recommendations for the remaining labeling tasks.

\subsection{User Experience}
\begin{figure}[h]
  \centering
  \frame{\includegraphics[width=0.9\linewidth]{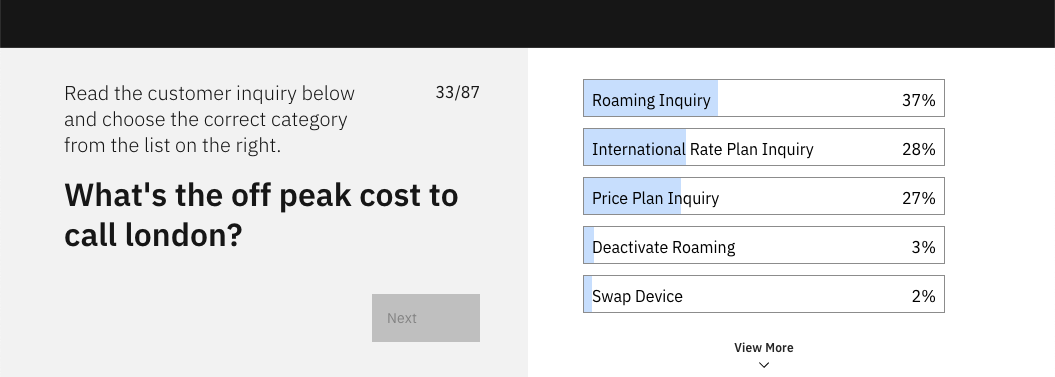}}
  \caption{The assisted labeling interface.}\label{fig:assisted_overview}
  \Description{The assisted labeling interface.}
\end{figure}

AI assistance is presented to the labeler via a visual system of ordering, annotating, and reducing the visible label set as shown in Figure ~\ref{fig:assisted_overview}. The list of labels is ordered based on the AI assistance prediction probabilities. The ordering is explained by overlaying probability information (blue bars) which serves as an indication of the assistant's confidence. 

\begin{figure}[h]
  \centering
  \frame{\includegraphics[width=0.9\linewidth]{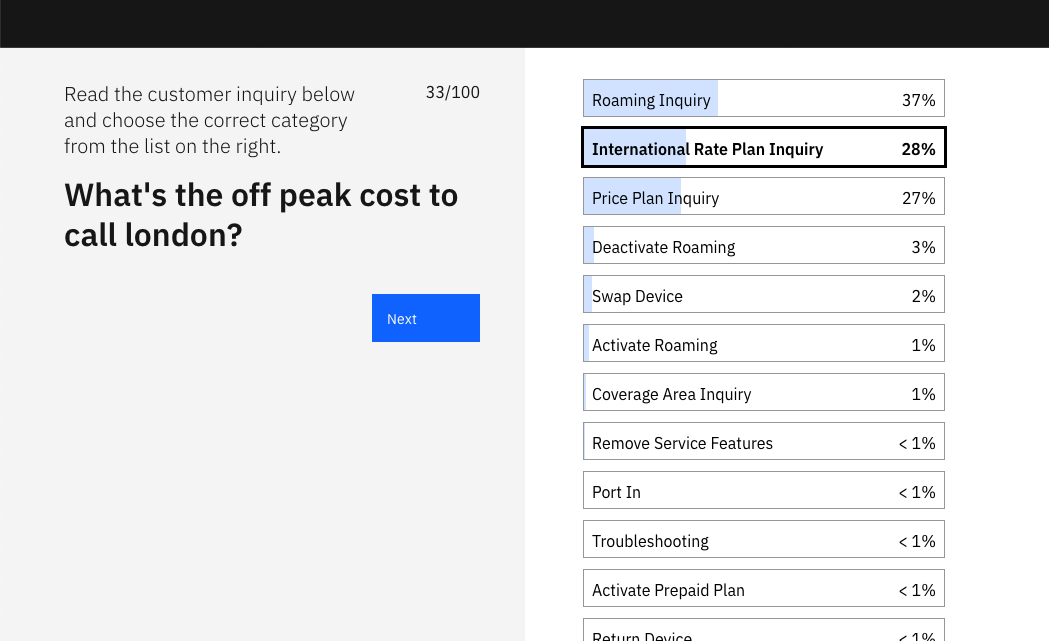}}
  \caption{The assisted labeling interface after the user clicked ``View More.''}\label{fig:assisted_overview_expanded}
  \Description{The assisted labeling interface.}
\end{figure}

The list of labels is initially limited to the top five
with the highest confidence scores. Limiting the label display to the top five reduces the complexity of the label decision; however, the user is free to expand the full list of labels using the ``View More'' option, as illustrated in Figure  \ref{fig:assisted_overview_expanded}.
We chose five based on an inspection of the probability and accuracy values of the AI models. As shown in Figure~\ref{fig:assisted_overview_expanded}, in most cases, probability values drop below one percent past the top five, rendering additional labels improbable from the perspective of the AI models. Also, in most cases, the true label of the example is among the top five. The label distribution is always shown rounded to whole percentages.
A score less than one but greater than zero is presented as ``<1\%'' so that all categories are possible answers; none has zero probability.

\section{Study Design}\label{sec:study}
We ran a between-subjects experiment to investigate whether our AI labeling assistance system improves labeling accuracy and speed, particularly in the context of a large label set. The study was designed to compare how labelers perform with and without AI assistance. Furthermore, the study considers two variants of AI assistance, weak and strong, to understand if improving the accuracy of the assistance also improves human performance.
We next describe the details of our study: the labeling task, conditions, methodology, our participants, materials, and the data collected.   



\subsection{Labeling Task \& Dataset}\label{subsec:task}
Participants were asked to examine and label a total of 100 (21 training + 79  experimental) short-text customer inquiries, within 30 minutes. The topic of the inquiries was consumer telecommunication service, and each inquiry had one `ground truth' label. There was a choice of 21 unique labels, each corresponding to a customer \textit{intent}. For example, the inquiry ``Why does my call drop every time I drive to work?'' is correctly labeled with the intent ``Network Complaints,'' meaning that the customer is unhappy about some aspect of the network and their intent is to complain. Table ~\ref{tab:labels} shows the 21 labels.  

We chose the dataset due to its general domain accessibility for non-specialists, its realistic use case, and because of the number of labels included. Our goal was to focus on larger label spaces to create a more challenging decision space. Considering work related to working memory capacity \cite{miller1956magicalnumber,cowan2010magicalmysteryfour}, feasibility of visual presentation, and availability of appropriate data, we deemed 21 labels to be sufficiently `large' for our purposes.

\begin{table}
  \small
  \caption{The 21 Study labels}
  \label{tab:labels}
  \begin{tabular}{l}
    \toprule
    Label\\
    \midrule
    Activate Device\\
    Activate Prepaid Plan\\
    Activate Roaming\
    Add Insurance\\
    Add Service Features\\
    Change Price Plan\\
    Coverage Area Inquiry\\
    Deactivate Prepaid Plan\\
    Deactivate Roaming\\
    Device Upgrade Eligibility\\
    International Rate Plan Inquiry\\
    Network Complaints\\
    Network Unlock\\
    Port In\\
    Price Plan Inquiry\\
    Recharge SIM\\
    Remove Service Features\\
    Return Device\\
    Roaming Inquiry\\
    Swap Device\\
    Troubleshooting\\
\end{tabular}
\end{table}

\subsection{Conditions}\label{subsec:conditions}
\begin{figure}[h]
  \centering
  \frame{\includegraphics[width=0.9\linewidth]{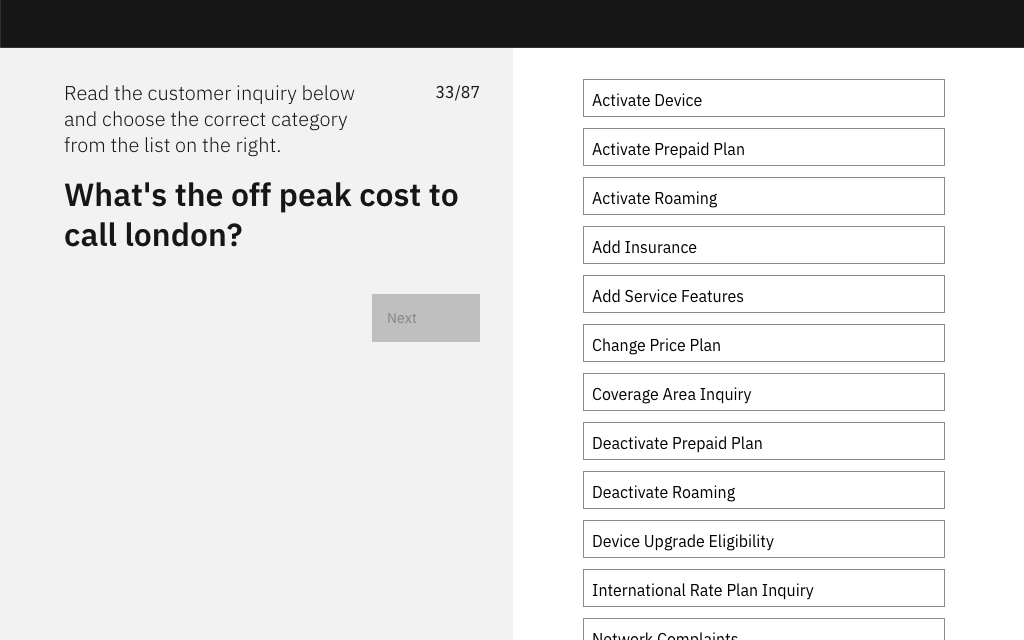}}
  \caption{The unassisted labeling interface used by the baseline group.}
  \Description{The unassisted labeling interface used in the baseline condition.}
  \label{fig:unassisted_interface}
\end{figure}

The study was organized into three distinct labeling conditions, spanning two interfaces. Subjects in the baseline \textit{unassisted} condition performed the labeling task with no AI assistance using the labeling interface in Figure ~\ref{fig:unassisted_interface}. The baseline interface presented all available labels in alphabetical order. 

Subjects in the \textit{weak assistance} condition were assisted by a predictive model that was minimally trained to predict labels, while those in the \textit{strong assistance} condition were assisted by a model with more training that provided more accurate predictions. Including both weak and strong assistance enabled us to analyze how performance of the assistant affects performance of the human labeler. 
Participants in both assisted conditions performed the study task using the assisted interface shown in Figure ~\ref{fig:assisted_overview} and ~\ref{fig:assisted_overview_expanded}, and the predictive performance of the assistant differentiated the conditions.

To implement the weak and strong assistance conditions with differing degrees of predictive performance, we varied the amount of labeled data input to the label-prediction algorithm and predicted two snapshot label distributions over the study dataset. 
Table \ref{tab:assistant_performance} describes the amount of labeled data, the model's accuracy on the study dataset, and top five accuracy---the percentage of examples for which the true label is included in the top five---for the two conditions.

\begin{table}
  \small
  \caption{AI-assistance performance details on the set of 79 study examples.}
  \label{tab:assistant_performance}
  \begin{tabular}{llll}
    \toprule
    Assistance Condition & \# Labeled & Acc. & Top 5 Acc.\\
    \midrule
    Weak    & 21 & 0.44 & 0.86 \\
    Strong  & 221 & 0.59 & 0.94 \\
    \bottomrule
\end{tabular}
\end{table}

\subsection{Methodology}
We prepared the participants for the study using an interactive training session, which was completed first and presented using the appropriate interface for the condition (baseline or assisted). During training each participant labeled 21 examples, one from each of the possible labels in the dataset. The examples used for training were chosen to be the most typical examples for their class/label. We selected these examples algorithmically by calculating the set of examples closest to the centroid (class mean vector) of each class. Participants were actively reinforced during the training phase. Upon selecting the correct label the participant was presented with a message indicating that the selection was correct, along with the corresponding label guideline. An incorrect selection resulted in a message indicating such, along with the correct label and its corresponding guideline. Participants were also reminded that label guidelines were available via contextual tooltips throughout the study. 

After completing the training phase, participants moved on to the experimental phase. They were asked to label the remaining 79 examples, one at a time. Unlike training, the participants did not receive feedback on whether the selected label was correct; however, the tooltips with label guidelines were available. 
The order of the 79 examples (which constituted the experimental portion of the study data) was random, and the order was reversed for one half of the participants exposed to each condition, to counteract any potential order effects.

\subsection{Participants}
A particular concern of the study was to create a realistic data-labeling scenario wherein the labelers were sufficiently motivated to perform the labeling task to the best of their abilities. We recruited our study participants via the Amazon Mechanical Turk platform and specifically requested `Master Workers' who have a track record of sincere effort. To motivate workers we advertised a monetary bonus linked specifically to labeling accuracy. Each participant earned a 2\textcent \ bonus per example correctly labeled in addition to the flat payment of \$4.25 (chosen to be above the current US hourly minimum wage) for completing the 30-minute study. A maximum bonus of \$1.58 could be earned on the task. 

Most of our participants had significant experience on Mechanical Turk: 43 had more than 3 years of experience, 7 had 2-3 years, and 4 had 1-2 years. Of our participants, 50 primarily spoke English, while 2 primarily spoke Hindi and 1 primarily spoke Tamil. All participants had at least a high school education, and 41 participants had formal education beyond high school. While most participants (30) had heard about AI, 13 closely followed AI-related news, 8 had AI related work experience, and 3 had extensive experience in AI research or development. 

\subsection{Materials}
The dataset used in the study contained a total of 420 examples, labeled by experts with experience designing customer assistance chatbots. The data was balanced among the 21 classes (labels) with some minor variance. We encoded each example in the dataset to a 512-dimensional vector using the Universal Sentence Encoder \cite{Cer2018}. 
To create a high-quality and consistent training experience for the participants to get familiar with the labeling problem, we calculated the `centroid' or mean vector for each class, and selected a set of 21 examples closest to the centroids as training data. We then randomly selected and removed a set of 79 of the remaining examples to be labeled by the participants during the study, ensuring the set was balanced across the 21 labels. This left us with 320 examples to implement the assistance conditions. 

\subsubsection{Implementing Weak Assistance} For the weak assistance condition, we selected 21 labeled examples from the 320 remaining examples using the same nearest-to-centroid approach. We then applied label spreading \footnote{We used Scikit Learn \cite{scikit-learn}'s implementation of label spreading (kernel=``knn'', n\_neighbors=10) for both weak and strong assistance.} to propagate labels from the 21 examples into the remaining 299 examples and the study data (79 examples) combined. This models a realistic scenario where label information is diffused from an initial labeled dataset into a larger unlabeled dataset. We used the predicted label distributions on the study data examples to create the weak assistance condition.  

\subsubsection{Implementing Strong Assistance} For the strong assistance condition we simply expanded the labeled data iteratively using a low-margin active learning selector. After 20 active learning iterations (adding 10 labeled examples per iteration) we again applied the label spreading algorithm to generate label distributions over the study dataset. These distributions were used in the strong assistance condition. 

\subsection{Data Collected}
The following information was collected from each participant during the study.

\subsubsection{Selected Labels} For each example presented we stored the label selected by the participants. We were then able to compare the participant's selection to the ground truth label to determine individual participant labeling accuracy. 


\subsubsection{Labeling Duration} The amount of time spent labeling was calculated as the duration between the labeler being presented with the example  and when the participant clicked the button to submit the label and transition to the next example. This approach removed any variation due to differences in page loading times. The labeling durations were also winsorized with a 5\% lower limit and a 95\% upper limit in order to reduce the the effect of spurious outliers, such as when a labeler would take a break or need to attend to some unrelated matter while performing their task. Winsorizaton was applied per participant.

\subsubsection{``View More'' events} We recorded all use of the ``View More'' function, which was available to the assisted conditions only. This event indicates that the participant expanded the visible subset of labels to see the full list. This would provide a valuable insight into how users interacted with the decision space reduction feature.

\subsubsection{Questionnaire}
Participants also completed a post-experiment questionnaire, which requested feedback about the labeling experience in general, and perceptions of the weak and strong AI if conditions allowed.
\section{Results}\label{sec:results}

A total of 54 participants successfully completed the experiment and an average bonus of \$1.20 was earned. Of the 54, \textbf{19} participants completed the baseline condition, \textbf{18} completed the weak assistance condition, and \textbf{17} completed the strong assistance condition. 


\subsection{Labeling Accuracy}
\begin{figure}[h]
  \centering
  \includegraphics[width=0.9\linewidth]{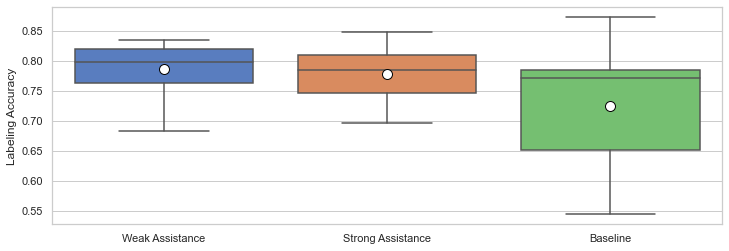}
  \caption{Mean labeling accuracy across the three study conditions. The mean value is represented by the white circle.}
  \Description{Mean labeling accuracy across the three study conditions.}
  \label{fig:accuracy}
\end{figure}

We performed a one-way ANOVA to evaluate the accuracy differences between the three study conditions: baseline, weak assistance, and strong assistance.  Data is presented as \textit{mean} $\pm$ \textit{standard deviation}. Accuracy was 
significantly different between the three conditions, F\textsubscript{2, 51} = 4.18, p < 0.02. Accuracy decreased in the baseline condition (0.72 $\pm$ 0.10) compared to the weak assistance condition (0.79 $\pm$ 0.05) and the strong assistance condition (0.78 $\pm$ 0.05), as shown in Figure ~\ref{fig:accuracy}. Tukey post-hoc analyses revealed that the increase from baseline to weak assistance was statistically significant (p < 0.03). Participants were more accurate in the weak assistance condition than in the baseline condition. No other condition differences were statistically significant. The accuracy difference between the baseline condition and strong assistance condition was not statistically significant (p < 0.07), but survey results suggest that strong assistance positively affected labeler accuracy.

In the post-study survey participants rated the helpfulness and accuracy of the label recommendations highly in both the weak and strong assistance conditions. Across assisted conditions, 91\% of participants rated the assistance as helpful. Of the three participants who did not find the assistance helpful, two were in the weak assistance condition. Fewer participants rated the suggested labels as accurate than helpful, with 80\% of participants agreeing that the assistance was accurate, 9\% neutral, and 11\% in disagreement. Using the Fisher test, we did not find significant differences between the strong and weak assistance conditions in how participants rated the helpfulness or accuracy of the assistance. Overall, participants often commented that the suggested labels were helpful \textit{"when they were right"}. 
\begin{quote}
\textit{"The ones where the recommendation was correct were exciting because it saved me time."} - P9-WA (weak assistance)
\end{quote}
These results support our hypothesis \textbf{H1} since our findings show that AI assistance increases accuracy in labeling tasks. 
A further breakdown of assisted labeling accuracy, in particular considering the use of the ``View More'' function, revealed the following. In general participants in the weak assistant condition used the ``View More'' feature in 29\% (28.8 $\pm$ 0.8) of labeling tasks, and participants in the strong assistant condition used the feature in 17\% (16.6 $\pm$ 0.7) of their tasks. This is consistent with the performance differences between the AI conditions, namely that better predictive performance would account for reduced necessity of the ``View More'' feature. In the cases where the view more option was used participants in the weak assistance condition achieved 66\% accuracy (65.9 $\pm$ 0.1), while those in the strong assistance condition achieved 60\% accuracy (60.1 $\pm$ 1.2). Alternatively when the view more option was \textbf{not} used participants in the weak assistance condition achieved 84\% accuracy (83.4 $\pm$ 0.4), while those in the strong assistance condition achieved 81\% accuracy (81.2 $\pm$ 0.4).



\subsection{Labeling Speed}
A general linear mixed-effects model (a.k.a. mixed model) analysis of variance was used to analyze labeling duration. The assistance condition and whether the ``View More'' option was selected were modeled as fixed effects, while trials were nested within each subject and modeled as a random effect.  Mixed models are more robust since they consider the output of each individual trial and also are more robust to missing data and unbalanced designs \cite{krueger2004comparison}.

\begin{figure}[h]
  \centering
  \includegraphics[width=0.9\linewidth]{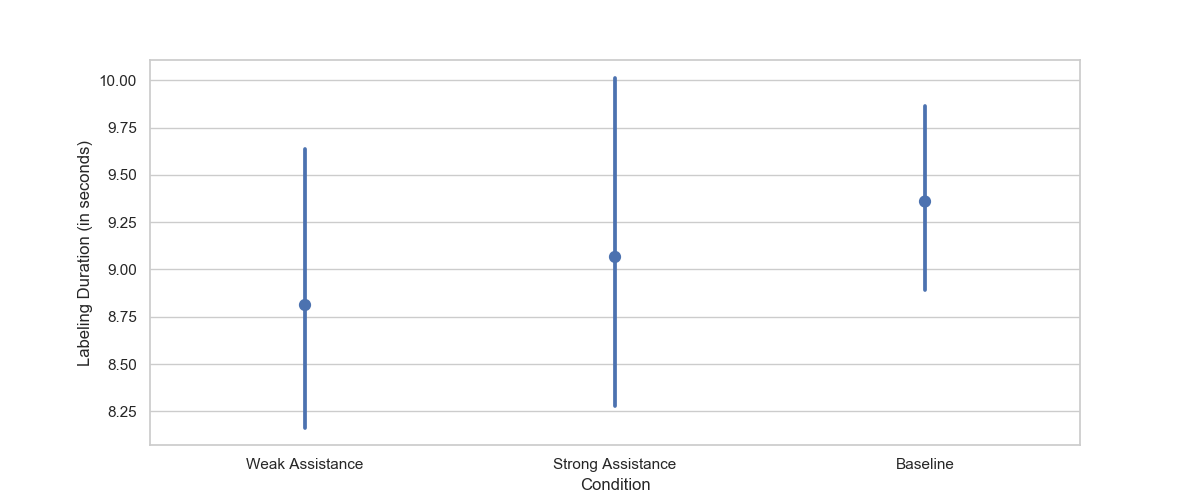}
  \caption{A plot of average duration of labeling time per trial for various assistance conditions.}
  \Description{graph}
  \label{fig:dur}
\end{figure}

There was a significant effect of assistance condition (baseline, weak assistance, strong assistance) on overall labeling duration (measured in seconds) F\textsubscript{2,58.2}=5.71, p < .001. Tukey post hoc tests showed that users spent less time on each trial in the weak assistance condition (8.81 $\pm$ 14.6) than baseline condition (9.36 $\pm$ 9.58), p < 0.004, and spent less time on each trial in the strong assistance condition (9.07 $\pm$ 16.5) than the baseline condition (9.36 $\pm$ 9.58), p < 0.04, as shown in Figure \ref{fig:dur}. 
There was also a significant effect of ``View More'' selection on duration (``View More'' selected, ``View More'' not selected) F\textsubscript{1,3730.3} = 202.59, p < 0.001; the trials in which participants selected the ``View More'' button (15.97 $\pm$ 15.17) were longer in duration than those who did not (7.89$\pm$ 13.12), regardless of the AI assistance condition. 
Additionally, when participants used the ``View More'' feature in the assistance conditions (strong (18.38 $\pm$ 22.50), and weak (14.65 $\pm$8.69)), they took longer to complete the labeling task than participants in the baseline condition (9.36 $\pm$ 9.58). We attribute this result to the fixed alphabetical order of labels in the unassisted condition, which was more efficient to use compared to the confidence ordered label list used in the assistance conditions, for more discussion on this see \ref{section:ordering_labels_discussion} 

 To better understand user behavior and the outcome of the labeling task, we must consider the user interface interaction effects of the use of ``View more'' and the condition. 
 We found a significant interaction of \textit{View More Selection} x \textit{AI Assistance Condition} on duration F\textsubscript{1,4253.5}=6.5. p < 0.05. Tukey post-hoc analysis revealed significant differences when ``View More'' was not selected between the baseline condition (9.36 $\pm$ 9.58) and the weak assistance condition (6.45 $\pm$ 15.79)   p\textless 0.05, but not between the baseline condition and the strong assistance condition, (7.21 $\pm$ 14.33) p=0.15. When users did not select the ``View More'' button, they completed the labeling task more quickly in the 
weak
assistance condition. 
When the ``View More'' button was selected, there were significant differences between the 
strong assistance condition (18.4 $\pm$ 22.5)  and the 
weak
assistance condition (14.65 $\pm$8.69)  p\textless 0.05. 
This result is likely related to the reduced usage of the ``View More'' feature (learning effects) and the more conservative confidence distributions (See section \ref{conf_pred}) exhibited by the strong assistance condition. 

\begin{figure}[t]
  \centering
  \includegraphics[width=0.9\linewidth]{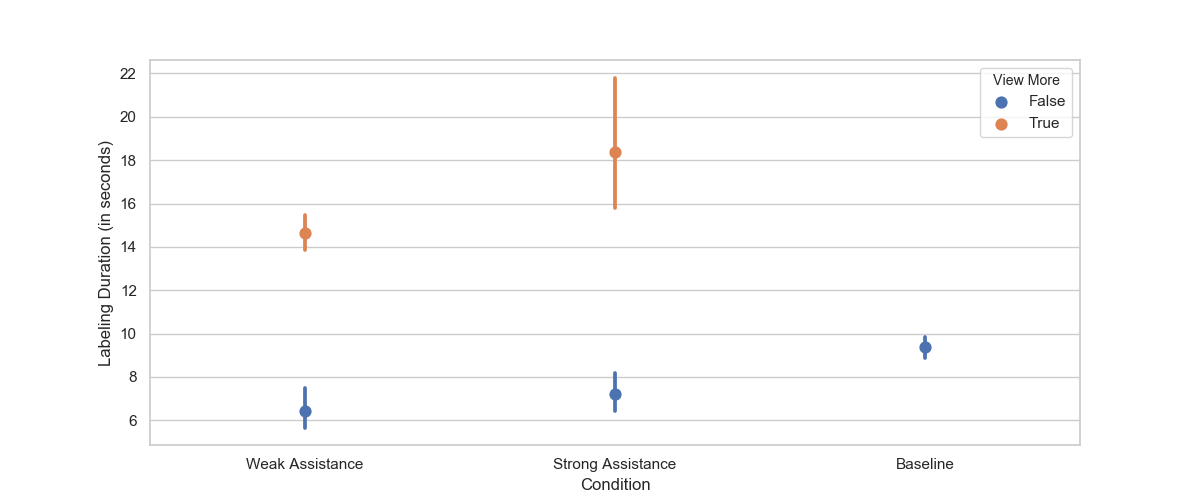}
  \caption{A plot of average duration of labeling time per trial for various assistance conditions. Hues represent whether ``View More'' button was selected.}
  \Description{graph}
  \label{fig:duration}
\end{figure}

Figure \ref{fig:duration} shows the durations for each of the conditions. Our results support \textbf{H2}, showing that AI assistance makes labeling tasks more efficient, by decreasing the amount of time users take per task, especially when the labeler finds the label in the reduced set of recommendations provided by the assistant and does not need to use the ``View More'' button. 


Our survey results provide further insight into the approach of reducing the initial label set to the top 5 most probable labels, and reordering of the label list by most probable. All participants in the baseline condition agreed that it was easy to find the label they were looking for. Most participants with labeling assistance (86\%) also found it easy to find the label they needed in their original five recommended labels. However, when participants with AI assistance found it necessary to ``View More'' to see the full set of labels, only 49\% of participants found it easy to find the label that they were looking for. When asked to describe their process looking through the long list of labels, participants wrote:
\begin{quote}
\textit{"If it wasn't in the top five, it was a bit annoying.  I just used the search feature in my browser (ctrl+f) to find what I wanted in the long list."} P23-WA
\end{quote}
\begin{quote}
\textit{"I had to look through them [the labels] more closely/taking more time to find one that I thought fit."} P31-SA (strong assistance)
\end{quote}
Participants cited the ``View More,'' the long length, and lack of organization of labels as features they liked the least.


\subsection{Variation in Predictive Accuracy}



Surprisingly, a comparison of labeler accuracy between the two assistance groups shows that improving the predictive performance of the AI did not lead to improved labeling accuracy. The accuracy difference between the weak assistance condition (0.79 $\pm$ 0.05) and the strong assistance condition (0.78 $\pm$ 0.05) was not statistically significant. Furthermore, the weak assistance group  used the ``View More'' feature significantly more often than the strong assistance group F(12.3) = 202.59, p \textless 0.001.
Recall that weak assistance had inferior model accuracy when compared to strong assistance, and labelers need to search for labels outside of the recommended set more often. In light of these results we reject \textbf{H3}. Participants were not more accurate when their labeling tasks were assisted by a more accurate predictive model. 

With any system that is automated, there is a risk that the user will overly rely on the system and relax their own perceptual abilities when making a decision \cite{heer2019agency}. This was a concern for us when designing the study. An overreliant labeler might simply accept the top recommended label without questioning the correctness. 
However, our results suggest that labelers were liable to make their own determination of correctness regardless of the quality of the assistant's recommendation. We analyzed the level of agreement between our participants and the AI predictions, and found that in the weak assistance condition participants agreed with the top predicted label in 45\% of tasks (45.0 $\pm$ 0.4) and in the strong assistance condition participants agreed with the top predicted label in 64\% (64.0 $\pm$ 0.5) of the tasks. This finding is consistent with prior work by Yu, Kun, et al. \cite{yu2016trust} which established that AI-assisted users were able to perceive the accuracy of an AI decision support system and adjust their trust accordingly. Sun et al. \cite{sun2017label} observed a similar pattern when studying the effectiveness of visual representations of learning during interactive labeling tasks. 



Participants' survey responses often reflected their efforts to find the correct response rather than merely relying on the recommended labels. Most participants seemed to consider the recommendations critically, rather than assuming that they were necessarily correct, such as P13:

\begin{quote}
\textit{The recommendations by the AI assistant gave a good `base' to go off of but was not full accurate enough.} P13-WA
\end{quote}

\section{Discussion and Design Recommendations}
\subsection{Ordering of Labels} \label{section:ordering_labels_discussion}
\textbf{The inclusion of predictive recommendations should compliment the human tendency to learn and exploit spatial knowledge in a consistent information space}.
In some of our open-ended responses in the survey following the labeling task, users complained about the ordering of labels changing.  In the assisted conditions the labels are ordered based on the confidence of the label predictions (both before and after clicking ``View More''), and a consistent spatial/alphabetical label order was never materialized. Some participants noted this in the survey, wishing it was alphabetical or ordered in another useful way. When asked about finding labels using ``View More'', participants wrote:
\begin{quote}
    \textit{"They didn't seem to be ordered very well. It should have been alphabetical."} P23-WA
\end{quote} 
\begin{quote}
\textit{"They weren't grouped in a way that made sense to me. They seemed to be randomly listed."} P26-WA
\end{quote}

Prior work has shown that users quickly form and rely on spatial knowledge in menu systems \cite{cockburn2007predictive}, that brief menus can lead to faster performance and learning or over-learning a menu-structure over time can provide performance enhancements ~\cite{ahlstrom2010s}. Watts-Perotti and Woods \cite{watts1999experienced} explain that when information is \textit{spatially dedicated} (kept in a consistent place within a display system) the location itself serves as a memory aid for users who become familiar with the structure. In our assisted labeling scenario, when a user is able to find the desired label in the reduced decision space they gain a significant efficiency advantage by working with a smaller set of relevant options. However when the user cannot find the correct label in the reduced set they need to expand the full set of labels and search for the label they want. We suspect that searching an inconsistently ordered set of items is far less efficient than relying on spatial knowledge learned from a consistent display space. As such further labeling efficiency might be achievable by providing an alphabetically ordered list of labels, annotated with model predictions, alongside the subset of recommended labels (Top 5 ordered by prediction confidence). Alternatively the labeler might be allowed to switch between probabilistic or alphabetical orderings of the label set on demand. 
These solutions would eliminate the mechanical task of opening the full set of labels (``View More'' button) and also allow the labeler to fall back on their spatial knowledge in the event that a satisfactory label prediction was not found.

\subsection{Confidence of Label Predictions } \label{conf_pred}
\textbf{Future designs should carefully consider the presentation of model confidence and set user expectations accordingly.} Prior work shows that when user expectations are high or do not match the level that the system is capable of delivering, users are less satisfied with their experience \cite{cheng2019explaining,wang2019designing,wiegand2019drive}. A participant noted in the survey that the system presented recommendations with high confidence that seemed incorrect:
\begin{quote}
\textit{"Sometimes it was clearly way off the mark and had very high confidence."} P10-SA 
\end{quote} 
Surprisingly, labelers assisted by the weak AI performed slightly better than those assisted by the strong AI, even through the strong AI demonstrated considerably better predictive performance on the labeling task. An analysis of the predictions provided by both AI assistants indicated that the weak AI was more \textit{certain} in its predictions than the strong AI. We define certainty as the inverse scaled entropy of the predictions provided by the assistant over the course of the labeling task. In particular the weak AI was prone to highly confident correct and incorrect predictions, while the strong AI exhibited a more conservative distribution of confidence scores across its predictions.

\begin{figure}[t]
  \centering
  \includegraphics[width=0.75\linewidth]{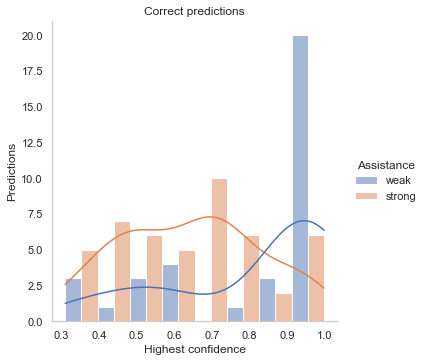}
  \caption{Top confidence of \textbf{correct} label predictions. The X axis represents confidence values on the unit interval. The Y axis represents the number of predictions. The weak assistant tended to produce more highly confident predictions. }
  \Description{graph}
  \label{fig:top_pred_conf_correct}
\end{figure}

\begin{figure}[t]
  \centering
  \includegraphics[width=0.75\linewidth]{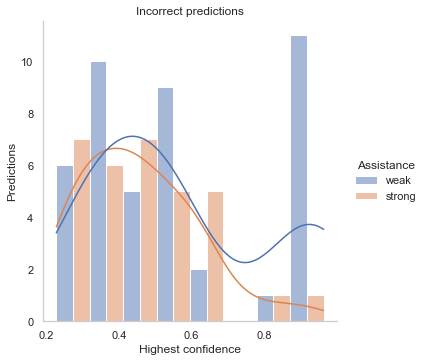}
  \caption{Top confidence of \textbf{incorrect} label predictions. The X axis represents confidence values on the unit interval. The Y axis represents the number of predictions. The weak assistant tended to produce more highly confident predictions. }
  \Description{graph}
  \label{fig:top_pred_conf_incorrect}
\end{figure}



A deeper comparison of labeling trials showed that in the cases where the assistant made an incorrect prediction, labelers agreed with the weak AI in 11\%  (11.1 $\pm$ 5.6) of those trials, but labelers agreed with the strong AI in 26\% (26.4 $\pm$ 11.3) of the trials, an unexpected difference between the two conditions. It's possible that the confident but often incorrect weak AI fostered a greater degree of distrust on behalf of the labelers, who were then more likely to scrutinize its predictions. On the other hand, the strong AI, being more often correct and exhibiting a more appropriate confidence range, may have caused labelers to more often trust incorrect predictions. Overall the labelers using the weak assistant performed slightly better, perhaps due to having a little less trust in the predictions of the assistant. See figures \ref{fig:top_pred_conf_correct} and \ref{fig:top_pred_conf_incorrect} which show the distributions of top confidence scores for correct and incorrect prediction across the two assistance conditions. Note the skew towards highly confident predictions for the weak assistance condition.

The effect of confidence scores on labeler performance needs further study. Characteristics like the certainty of predictions and confidence presentation may have unexpected consequences on the decision making process.

\subsection{AI Assistance in low information environments}
\textbf{AI assistance appears to have value in the early stages of labeling activities, but having a representative set of training examples is vital}.
AI assistance can improve labeling performance even with very little labeled data. Our weak assistant only had 21 labeled examples in order to make predictions. However provision of representative examples to train the labeling assistance model is a fundamental concern. A skewed or incomplete set of training examples may cause labeling errors which compound over time, as mislabeled examples from previous labeling iterations are rolled into the assistants training data. It would be pragmatic to provide assistance only when a representative set of labeled examples is available. In our experiment we selected one labeled example per label as training for the weak assistant, but this requirement may change depending on the particulars of the dataset and the labeling task under consideration.

\section{Limitations}

This paper reflects one particular context. We studied a single task in this paper; future work should extend to additional types of tasks with varying degrees of difficulty. Also, we recruited participants from Mechanical Turk, which is known to have intersectional gender imbalances in different countries and cultures \cite{difallah2018demographics}. Mechanical Turk workers are not subject matter experts and results may differ for labeling tasks that require domain expertise. We do believe the results of this study can be leveraged in future work exploring AI assistance for datasets that require subject matter expertise.  

\section{Conclusion}\label{sec:conclusion}

In this paper we studied the impact of AI assistance on human data labeling performance. In particular we focused on a labeling task with a larger label set than had previously been studied. 

Our results indicate that using a predictive model as an `AI assistant' can improve labeling accuracy. We found that varying the predictive performance of the AI did not significantly impact the accuracy of human labelers. Participants in the study did not over-rely on the predictions of the labeling assistant, and it seems that the accuracy of human labeling can be improved with relatively weak AI support.


We also studied the effects of AI assistance on labeling speed. We found that reducing the labeler's ``decision space'' by showing only the top 5 most confidently predicted labels, improves labeling speed, particularly when the labeler finds the desired label in the filtered subset. When labelers need to expand the full label set using the ``View More'' function, the speed improvements are reversed. Future work will involve understanding how to  optimally present label predictions without negatively affecting the labelers ability to efficienty navigate and select from the full set of labels. 






In the course of the study, and during subsequent analysis of the results, we noticed subtle effects of prediction confidence on labeling behavior. Further study is required to understand such phenomena. We also acknowledge that to understand the full potential of AI assistance for data labeling, a broader range of studies with varying degrees of task difficulty and label set size would be worthwhile.

\bibliographystyle{ACM-Reference-Format}
\bibliography{ai_assistance}










\end{document}